\documentclass[floatfix,aps,twocolumn,nopacs,preprintnumbers,amsmath,amssymb]{revtex4-1}

\usepackage[applemac]{inputenc}
\usepackage[T1]{fontenc}
\usepackage{amsfonts, amssymb, amsmath, dsfont}
\usepackage{graphicx}
\usepackage{lipsum}
\usepackage{float}
\usepackage{ulem}
\usepackage{color}

\newcommand{\bcalM}{\boldsymbol{\mathcal{M}}}
\newcommand{\bcalP}{\boldsymbol{\mathcal{P}}}

\newcommand{\bF}{\mathbf{F}}
\newcommand{\bbf}{\mathbf{f}}
\newcommand{\PI}{\mathbf{\Pi}}
\newcommand{\bLamb}{\mathbf{\Lambda}}
\newcommand{\bPhi}{\mathbf{\Phi}}

\newcommand{\bE}{\mathbf{E}}

\newcommand{\bH}{\mathbf{H}}

\newcommand{\bT}{\mathbf{T}}

\newcommand{\br}{\mathbf{r}}

\newcommand{\bp}{\mathbf{p}}
\newcommand{\bbm}{\mathbf{m}}

\newcommand{\tk}{\tilde{k}}
\newcommand{\tq}{\tilde{q}}

\def\d{\mathrm{d}}
\def\m{\mathrm{m}}
\def\r{\mathrm{r}}
\def\eps{\varepsilon}
\def\Re{\mathrm{Re}}
\def\Im{\mathrm{Im}}

\def\dip{\mathrm{dip}}

\begin{document}

\title{Plasmonic lateral forces on chiral spheres}

\author{Antoine Canaguier-Durand}
\affiliation{ESPCI ParisTech, PSL Research University, CNRS, Institut Langevin, 1 rue Jussieu, F-75005, Paris, France.}
\author{Cyriaque Genet}
\affiliation{ISIS \& icFRC, University of Strasbourg and CNRS, 8 all\'{e}e Gaspard Monge, 67000 Strasbourg, France.}

\begin{abstract}
We show that the optical force exerted on a finite size chiral sphere by a surface plasmon mode has a component along a direction perpendicular to the plasmon linear momentum. We reveal how this chiral lateral force, pointing in opposite directions for opposite enantiomers, stems from an angular-to-linear crossed momentum transfer involving the plasmon transverse spin angular momentum density and mediated by the chirality of the sphere. Our multipolar approach allows us discussing the inclusion of the recoil term in the force on a small sphere taken in the dipolar limit and observing sign inversions of the lateral chiral force when the size of the sphere increases. 
\end{abstract}

\maketitle

\section*{Introduction}

The possibility to exert new types of forces on chiral objects by chiral optical fields has recently initiated a lot of interest \cite{canaguier2013mechanical,cameron2014discriminatory,cameron2014diffraction,bradshaw2014chiral,tkachenko2013spin,tkachenko2014optofluidic,tkachenko2014helicity,ding2014realization,wang2014lateral,chen2014tailoring,hayat2014lateral,canaguier2014chiral,alizadeh2015plasmonically,canaguier2015chiral}. Such forces have been observed at the level of micron-scaled chiral lipidic objects \cite{tkachenko2013spin,tkachenko2014optofluidic,tkachenko2014helicity} but the challenge is to exploit such forces at the nanoscale, with genuine chiral separation schemes at the molecular level, as proposed recently \cite{canaguier2013mechanical,cameron2014discriminatory,cameron2014diffraction,bradshaw2014chiral,wang2014lateral,chen2014tailoring,hayat2014lateral,canaguier2014chiral,alizadeh2015plasmonically}. This target, if proved viable with high throughputs, can obviously have an important technological impact given the critical importance of chiral separation processes in chemistry, in biology, and in the pharmaceutical industry. 

But chirality is a weak signature of matter: for most chiral organic molecules, the ratio between circular dichroism and absorption - the so-called anisotropy $g$ factor - is smaller than $10^{-3}$. As a consequence, chiral optical forces are {\it a priori} expected to be largely screened by usual (non-chiral) optical forces. When targeting the nanoscale, artificial objects that have been explicitly engineered so as to lead to maximized chiral strengths becomes particularly appealing. A few relevant examples can be found at the level of DNA-nanoparticle hybrids \cite{yan2012self} or  chiral golden nanopyramids \cite{mcpeak2014complex}. Such systems can display $g$ factors larger than $10^{-2}$. Simultaneously, it is important to realize that the limitations put by this unfavorable ratio can be evaded by developing specific optical schemes where the influence of chiral forces on the mechanical action is clearly identified, either by minimizing, if not merely canceling, non-chiral forces \cite{canaguier2013mechanical,ding2014realization} or by forcing the chiral-dependent component of the optical force acting on a different direction than the non-chiral action \cite{cameron2014discriminatory,cameron2014diffraction,wang2014lateral,hayat2014lateral}.

In this context, the possibility of a mechanical action performed by a light field perpendicularly to the incident photon momentum, so-called lateral force, is interesting. Lateral forces have been discussed lately on non-chiral \cite{bliokh2014extraordinary,rodriguez2015lateral} and chiral \cite{hayat2014lateral} dipoles as well as on chiral helices \cite{wang2014lateral}. From a physical point of view, they stem either from the self interaction of the chiral dipole moments \cite{wang2014lateral,rodriguez2015lateral}, or from a transverse spin momentum generated from a two-wave interference \cite{bekshaev2015transverse} or an evanescent wave \cite{hayat2014lateral}. In all cases, lateral forces originate from an interference between the electric and magnetic fields scattered by the dipole, i.e. from the so-called recoil term that appears when considering a dipole as a genuine physical system. 

In this article, we consider an original configuration where a finite size chiral sphere is immersed in a plasmonic near field and we predict a purely chiral optical force component in the lateral direction, with opposite directions for opposite enantiomers. Such a force can therefore be used for separating finite size chiral objects as a function of their enantiomeric forms. Our choice of configuration actually serves two purposes. First, we show that the transverse spin angular momentum density, characteristic of surface plasmon (SP) modes, is exactly what is needed to induce a lateral component in the optical forces via an angular-to-linear crossed momentum transfer mediated by the chirality of the scatterer. Secondly, we show that the use of a dipolar recoil term when using the dipole model as the small size limit for a nanoparticle can lead to inaccurate results, since this recoil term can be overtaken in magnitude by the multipolar response of the chiral object. 

Nevertheless, we show that Mie-based multipolar calculations for a finite size chiral sphere within our configuration confirm the existence of a lateral chiral force, just as predicted by the recoil term on a model chiral dipole. Interestingly, a quasi linear dependence of the lateral force with respect to the sphere radius is observed for large chiral spheres that displays optical rotation. For the case of a sphere with circular dichroism, we reveal that the progressive inclusion of multipoles in the lateral force evaluation can lead to a sign inversion of the force when the sphere radius increases. This sign inversion is clearly reminiscent of the pulling effect we described recently involving angular-to-linear crossed momentum transfers mediated by chirality \cite{canaguier2015chiral} and could have important consequences when discussing discriminatory protocols based on the lateral chiral force effect. 

In the first part of this article, we briefly gather the general expressions for the chiral and non-chiral optical forces on a nanosphere in the dipolar regime and discuss the inclusion of the recoil force. Then in a second section we apply these results to the case of a plasmonic field, in order to show that a lateral force appears in the recoil optical force with opposite directions for opposite enantiomeres. In a third part we finally present a multipolar calculation of the plasmonic force exerted on a chiral sphere, which enables to confirm the existence of this chiral lateral force and to discuss the validity of the recoil force in the small sphere limit.

\section{Optical forces in the dipolar regime}

The optical response of a nanoparticle whose dimensions are much smaller than the spatial variations of the surrounding harmonic electromagnetic field is usually described in the dipolar regime, by introducing electric and  magnetic dipolar moments $\bcalP=\Re[\bp_0(\br) e^{-\imath \omega t}]$ and $\bcalM=\Re[\bbm_0(\br) e^{-\imath \omega t}]$. In the case of a nanoparticle with chiral properties located in a fluid, the later moments depend on both the incident electric and magnetic fields as:
\begin{align}\label{def_dipole_alpha_beta_chi}
\left( \begin{array}{c} \bp_0 \\ \bbm_0 \end{array} \right) =  \left( \begin{array}{cc} \alpha \eps_\d &   \imath \chi \sqrt{\eps_\d \mu_\d}   \\ -\imath \chi   \sqrt{\eps_\d \mu_\d} & \beta \mu_\d \end{array} \right) \times  \left( \begin{array}{c}\bE_0 \\\bH_0 \end{array} \right)
\end{align}
where the electric, magnetic and mixed electric-magnetic dipole polarizabilities $(\alpha, \beta, \chi)$ have the dimension of a volume and $\eps_\d,\mu_\d$ are the dielectric permittivity and magnetic permeability of the surrounding fluid, respectively. For simplicity, we assume bi-isotropic response such that $(\alpha,\beta,\chi)$ are complex numbers. For a spherical nanoparticle, it can be shown that the dipolar polarizabilities can be expressed simply in the quasi-static limit ($\omega R/c \rightarrow 0$) from the parameters of the bulk material $(\eps,\mu, \kappa)$ constituting the nanosphere \cite{canaguier2015chiral}:
\begin{align}
\alpha &= 4\pi R^3 \frac{(\eps_\r-1)(\mu_\r+2) - \kappa^2}{(\eps_\r+2)(\mu_\r+2) - \kappa^2} \label{def_alpha} \\
\beta &= 4\pi R^3 \frac{(\mu_\r-1)(\eps_\r+2) - \kappa^2}{(\eps_\r+2)(\mu_\r+2) - \kappa^2} \label{def_beta} \\
\chi &= 12\pi R^3 \frac{ \kappa}{(\eps_\r+2)(\mu_\r+2) - \kappa^2} \label{def_chi}
\end{align}
where $\eps_\r=\eps/\eps_\d$ and $\mu_\r=\mu/\mu_\d$ are the relative permittivity and permeability, with respect to the surrounding fluid, and $R$ is the radius of the nanosphere. While $(\eps,\mu)$ give the refractive index $n$ of the nanosphere, the parameter $\kappa$ characterizes its chirality and corresponds to the effective refractive index difference $n_\pm =n\pm \kappa$ between left $(-)$ and right $(+)$ handed circularly polarized waves. The real part of $\kappa$ is thus associated with optical rotation and the imaginary part with circular dichroism. 

The time-averaged optical force applied to such a chiral dipole by a general harmonic field writes from the Lorentz force law as \cite{canaguier2013mechanical}:
\begin{align}\label{force_dipolar_start}
\bF_{\rm dip} &= \frac{1}{2} \Re\left[ \left( \bp_0 \cdot \nabla \right) \bE_0^* + \left( \bbm_0 \cdot \nabla \right) \bH_0^* \right. \nonumber \\
& ~ ~ ~ ~ ~ ~ \left. - \imath \omega \mu_\d \bp_0 \times \bH_0^* + \imath \omega \eps_\d \bbm_0 \times \bE_0^* \right]
\end{align}
where the optical fields are evaluated at the origin. We refer to the later quantity as \textit{strictly dipolar}, as it does not yet include terms stemming from self-interactions of the dipole, which will be the role of the additional recoil terms discussed below. When replacing in Eq.~(\ref{force_dipolar_start}) the dipolar moments by associated polarizabilities using Eq.~(\ref{def_dipole_alpha_beta_chi}), the force splits into additive non-chiral and chiral contributions, that depend only on $(\alpha,\beta)$ and $\chi$, respectively. Each contribution is again separable into reactive and dissipative components that respectively involve the real and imaginary parts of the $\alpha,\beta$ or $\chi$ polarizabilities:
\begin{align}
\bF^\mathrm{reac}_{\alpha,\beta} &= \Re[\alpha] \nabla W_E + \Re[\beta] \nabla W_H \label{F_achi_r} \\
 \bF^\mathrm{diss}_{\alpha,\beta} &=  \Im[\alpha] \frac{k_\d^2}{\omega} \PI_O^{(E)} +  \Im[\beta] \frac{k_\d^2}{\omega} \PI_O^{(H)}  \label{F_achi_d} \\
  \bF^\mathrm{reac}_\chi &= \Re[\chi] \frac{1}{k_\d} \nabla K  \label{F_chi_r} \\
 \bF^\mathrm{diss}_\chi &=  \Im[\chi] \frac{2k_\d}{\omega} \left( \bPhi - \frac{\nabla \times \PI }{2} \right) \label{F_chi_d} 
\end{align}
where $k_\d=n_\d \omega/c$ is the wavevector in the surrounding fluid whose refractive index is $n_\d=\sqrt{\eps_\d \mu_\d}/\sqrt{\eps_0 \mu_0}$. In these expressions, $W_E$ ($W_H$ ) is the energy density of the electric (magnetic) field respectively, $\PI_O^{(E)}$ ($\PI_O^{(H)}$) the electric (magnetic) orbital parts of the Poynting vector $\PI =  \Re \left[ \bE_0 \times \bH_0^* \right] / 2$. As discussed in \cite{canaguier2013mechanical}, $K=\Im \left[ \bE_0 \cdot \bH_0^*\right] k_d^2 / (2 \omega)$ is associated with the chirality density of the field and $\bPhi$ corresponds to the chirality flow, two quantities that obey a conservation equation similarly to the one relative to electromagnetic energy. The chirality flow $\bPhi$ is directly given by the electric and magnetic field ellipticities 
\begin{align}
\bPhi_E &=-\frac{1}{2} \Im\left[\bE_0 \times \bE_0^* \right]  \\
\bPhi_H &=-\frac{1}{2} \Im\left[\bH_0 \times \bH_0^* \right] \label{expression_PhiH}
\end{align}
with $\bPhi = \omega  (\varepsilon_d \bPhi_E + \mu_d \bPhi_H) / 2$.

Besides the dipolar Lorentz force law, one can add a recoil contribution to the force that stems from the self interaction of the dipole. While this term is important when considering the dipole as a genuine physical system to guarantee energy and momentum conservation \cite{nieto2015optical}, resorting to it is questionable when the dipolar model is used to represent a nanoparticle in the small-size limit. This will be discussed in the last section for finite size objects when confronting this recoil contribution to multipolar calculations. The recoil force involves both the electric and magnetic dipole moments and writes \cite{radescu2002exact,chaumet2009electromagnetic,nieto2010optical,chen2011optical,bekshaev2013subwavelength}:
\begin{align}\label{F_recoil_pm}
\bF_\mathrm{recoil} &= - \frac{\omega}{12 \pi} k_\d^3 \Re \left[ \bp_0 \times \bbm_0^*  \right] \nonumber \\
&= - \frac{k^4_\d}{6 \pi}   \left[ \left( \Re[\alpha \beta^*] + |\chi|^2 \right) \frac{n_\d \PI}{c} + \Im[\alpha \beta^*]  \frac{n_\d \bT}{c}  \right. \nonumber \\
& ~ ~ ~ ~ \left. \vphantom{ \frac{n_\d \PI}{c}} + \Re[\alpha \chi^*] \eps_\d \bPhi_E + \Re[\beta \chi^*] \mu_\d \bPhi_H \right]  
\end{align}
where one can see that this force scale with $R^6$ for a nanosphere, while the strict dipolar result $\bF_{\rm dip} \propto R^3$. Here $\bT=-\Im[\bE_0\times \bH_0^*]/2$ is sometimes called the ``imaginary'' Poynting vector \cite{bliokh2014extraordinary} as the imaginary part of the vector product is taken, while it is the real part in the case of the Poynting vector . 

This expression, consistent with the results of \cite{bliokh2014extraordinary,wang2014lateral,hayat2014lateral} on lateral forces, clearly reveals how field ellipticities can contribute to the optical force. Fundamentally, the flow of chirality is proportional to the spin angular momentum density $\bLamb_S$ with $\bLamb_S = \bPhi / k_d^2$ \cite{canaguier2015chiral}. Considering that for a transverse magnetic (TM) evanescent wave, the field ellipticity is reduced to the electric one, the third term in the right hand side of Eq. (\ref{F_recoil_pm}) shows how the chiral polarizability $\chi$ is able to transfer the spin angular momentum of the field into linear momentum for the particle in the form of a contribution to the recoil optical force. The dual symmetric picture obviously holds for a transverse electric (TE) evanescent wave. For both types of evanescent waves, the spin angular momentum turns out to be transverse, leading to a transverse component of the recoil force \cite{hayat2014lateral}.

\section{Plasmonic lateral force in the dipolar regime}

The same is true for an SP mode, as we now briefly detail by considering a planar harmonic plasmonic field launched and propagating along the $x$-axis on an interface between a fluid in the upper $(z>0)$-half space and a metal in the lower $(z<0)$-half space. The fluid (water) is characterized by a real dielectric function $\eps_\d$, while the metal is characterized by a complex dielectric function $\eps_\m$. For both material, we will take the permeability of vacuum ($\mu_\d=\mu_\m=\mu_0$). A nanosphere of radius $R$ has its center located in the fluid at a height $h$ above the interface, and is made of a chiral material with parameters $(\eps,\mu,\kappa)$. Taking the origin at the center of the sphere, the SP field in the fluid writes as:
\begin{align}\label{plasmonic_field2}
&\bE_0(x,y,z) =  E_0  e^{\imath kx} e^{\imath q (z+h)}   \left( \tq , 0 , -\tk \right)^t  \nonumber \\
&\bH_0(x,y,z) = H_0  e^{\imath k x} e^{\imath q (z+h)}  \left(  0, 1 , 0 \right)^t 
\end{align} 
where $H_0=\sqrt{\eps_\d/\mu_\d}E_0$, and the complex wavenumbers $k$ and $q$ of the plasmonic field in the longitudinal and vertical directions
\begin{align*}
&k=k'+\imath k'' = \frac{\omega n_\d}{c} \sqrt{\frac{\eps_\m}{\eps_\d + \eps_\m}}  & ~ ~  \tilde{k} = \tk' + \imath \tk'' =\frac{kc}{n_\d \omega}  \\
&q=q' + \imath q'' = \frac{\omega n_\d}{c} \sqrt{\frac{\eps_\d}{\eps_\d + \eps_\m}}  & ~ ~ \tilde{q} = \tq' + \imath \tq'' = \frac{qc}{n_\d \omega} 
\end{align*} 
are determined by the material parameter for the metal and the fluid. Let us stress that in the visible range and for most common metals (noble metals, e.g.), one has $k', k'', q''\geq0$ with $q'\leq0$.

For such an SP mode, the Poynting vector in the fluid 
\begin{align}\label{expression_PI}
\PI = \frac{1}{2} \Re \left[ \bE_0 \times \bH_0^* \right] = \frac{I_0}{2} e^{-2k''x-2q''(z+h)} \left( \begin{array}{c} \tk' \\ 0 \\ \tq' \end{array}\right) 
\end{align}
 is in the plane orthogonal to $\bH$, along the propagation direction but pointing downwards towards the interface. Here we have introduced for simplicity the field intensity $I_0=E_0 H_0^*$. As we consider a TM-transverse mode, only the electric field has a non-zero ellipticity:
\begin{align}
\bPhi_E &=- \sqrt{\frac{\mu_\d}{\eps_\d}} I_0 ~e^{-2k''x-2q''(z+h)} \left( \tk'\tq'' - \tk''\tq' \right) \left( \begin{array}{c} 0 \\ 1 \\ 0 \end{array}\right) \label{expression_PhiE}\\
\bPhi_H &=\mathbf{0}\label{expression_PhiH}
\end{align}
which yield the following orbital parts for the Poynting vector appearing in Eq.~(\ref{F_achi_d}):
\begin{align*}
&\PI_O^{(E)} = \PI - \frac{\nabla \times \bPhi_E}{2\omega \mu_\d} = \left( 1+ 2(\tk'')^2 + 2(\tq'')^2 \right) \PI \\
&\PI_O^{(H)} = \PI - \frac{\nabla \times \bPhi_H}{2\omega \eps_\d} =\PI ~ .
\end{align*}
Next, the energy densities  in Eq.~(\ref{F_achi_r}), carried by the electric and magnetic fields are, respectively: 
\begin{align*}
W_E  &= \frac{\eps_\d}{4} \| \bE_0 \|^2 \\
 &= \frac{n I_0}{4 c} e^{-2k''x-2q''(z+h)}  \left( 1+ 2(\tk'')^2 + 2(\tq'')^2 \right) \\
W_H  &= \frac{\mu_\d}{4} \| \bH_0 \|^2 = \frac{n I_0}{4c} e^{-2k''x-2q''(z+h)} ~ .
\end{align*} 
Because of the transversality, the electric and magnetic fields are orthogonal and the chirality density involved in Eq.~(\ref{F_chi_r}) is zero. Then, the flow of chirality is easily obtained from the ellipticities given in Eqs.~(\ref{expression_PhiE},\ref{expression_PhiH}):
\begin{align*}
K &= \frac{k_\d^2}{2\omega} \Im\left[ \bE_0 \cdot \bH_0^* \right] = 0 \\
\bPhi &= \frac{\omega}{2} \left(\eps_\d \bPhi_E + \mu_\d \bPhi_H \right) \\
&= - \frac{k_\d I_0}{2} e^{-2k''x-2q''(z+h)} \left( \tk'\tq'' - \tk''\tq' \right) \left( \begin{array}{c} 0 \\ 1 \\ 0 \end{array}\right) ~ .
\end{align*}
This corresponds exactly to the remarkable fact that plasmonic modes are characterized by a {\it transverse} spin angular momentum density \cite{bliokh2012transverse,canaguier2013force,canaguier2014transverse}. The importance of such a transverse density has been emphasized recently with various optical fields \cite{bliokh2012transverse,banzer2013photonic,canaguier2013force,canaguier2014transverse,neugebauer2014geometric,bekshaev2015transverse,neugebauer2015measuring,aiello2015transverse}. In the context of optical forces, it becomes the source for the lateral component of the recoil force exerted on a chiral dipole, as we now show.

When considering the optical force $\bF_\mathrm{dip}$ in the strict dipolar regime, such a plasmonic field does not yield any chiral force despite the non-zero flow of chirality. This comes from the fact that the curl of the Poynting vector exactly compensates this flow of chirality as $\bPhi - \nabla \times \PI  / 2 = \mathbf{0}$ -see \cite{canaguier2014chiral} of a detailed discussion. One is thus left non-chiral dipolar forces which explicitly writes as:
\begin{widetext}
\begin{align}\label{F_dip}
\frac{\bF_\dip}{n_\d I_0/c} = \frac{k_\d e^{-2q''h}}{2} \left\{ - \left[ \Re[\alpha] \left( 1+ 2(\tk'')^2 + 2(\tq'')^2 \right) + \Re[\beta] \right]  \left( \begin{array}{c} \tk'' \\ 0 \\ \tq'' \end{array}\right) + \left[ \Im[\alpha] \left( 1+ 2(\tk'')^2 + 2(\tq'')^2 \right) + \Im[\beta] \right]  \left( \begin{array}{c} \tk' \\ 0 \\ \tq' \end{array}\right) \right\}.
\end{align}
\end{widetext}

Such forces are fully contained in the $(x,z)$ plane showing that no lateral force in the $y$ direction is expected in the strict dipolar regime.

As discussed above, the recoil force however opens different perspectives. In the plasmonic case, we derive from Eq. (\ref{F_recoil_pm}) at the sphere position ($x=0, z=0$):
\begin{widetext}
\begin{align}\label{F_recoil}
\frac{\bF_\mathrm{recoil}}{n_\d I_0 / c} = \frac{k_\d^4 e^{-2q'' h}}{6\pi} \left[ - \frac{\Re[\alpha \beta^*] + |\chi|^2}{2}  \left( \begin{array}{c} \tk' \\ 0 \\ \tq' \end{array}\right) + \frac{\Im[\alpha \beta^*]}{2} \left( \begin{array}{c} \tk'' \\ 0 \\ \tq'' \end{array}\right) +  \Re[\alpha \chi^*]  \left( \tk'\tq'' - \tk''\tq' \right)  \left( \begin{array}{c} 0 \\ 1 \\ 0 \end{array}\right) \right].
\end{align}
\end{widetext}
While the first and second terms only give an additional contribution to the optical force in the $x$ and $z$ direction, the third term in contrast corresponds to a lateral force in the $y$ direction. This term is proportional to $\Re[\alpha \chi^*]$ and exactly corresponds to the transfer, mediated by the chirality of the dipole, of  optical transverse angular momentum into transverse linear momentum for the particle.  Such a lateral component in the recoil optical force generated by a plasmonic field is similar to what has been obtained recently with an evanescent mode excited with elliptical polarization in \cite{hayat2014lateral}. 

\section{The lateral force in the multipolar regime}

As it involves the product of momenta $ \bp_0 \times \bbm_0^* $, the recoil force $\bF_\mathrm{recoil}$ scales as $R^6$ for a small sphere, whereas the force obtained in the strict dipolar limit $\bF_\mathrm{dip}$ scales as $R^3$. This already indicates that for the case of a sphere made of a chiral medium, this additional force contribution is not rigorously a next-to-leading-order correction in the small size expansion. Indeed, the dipolar polarizabilities derived in Eqs.~(\ref{def_alpha}-\ref{def_chi}) are based on a first order expansion, which only includes terms that scale as $R^3$ while higher order multipolar terms, like the electric and magnetic quadrupoles would typically give contributions scaling with $R^5$. Such quadrupolar terms must therefore be included in the small-size expansion before taking into account the recoil force.

Hence in the case of a small nanosphere taken in the small-size limit, the result obtained with the additional recoil term derived for a generic $(\bp,\bbm)$ dipole should be checked with a more rigorous multipolar calculation, that would include all terms of successive orders in $R$ in the description of the field scattered by the sphere. As stressed above, this is the only way to check whether the lateral force is a genuine dynamical signature of chirality when considering real, finite size, scatterers. This is what we do in this section, using a generalized chiral Mie calculation that we presented earlier in \cite{canaguier2015chiral}. Our approach evaluates the optical force from the coefficients $A_{\ell,m}, B_{\ell,m}$ of the incident plasmonic field in spherical basis, which we derived in \cite{canaguier2014transverse}. In the following, we will use normalized forces $\bbf=\bF \times e^{2q^{''}h} /(n_\d I_0/c) $ for simplicity, and all numerical calculations are done for an incident plasmonic field with wavelength $\lambda = 594$ nm in vacuum.

\subsection{Forces on a non-dissipative dielectric sphere with optical rotation}

\begin{figure}[htbp]
\includegraphics[width=0.4\textwidth]{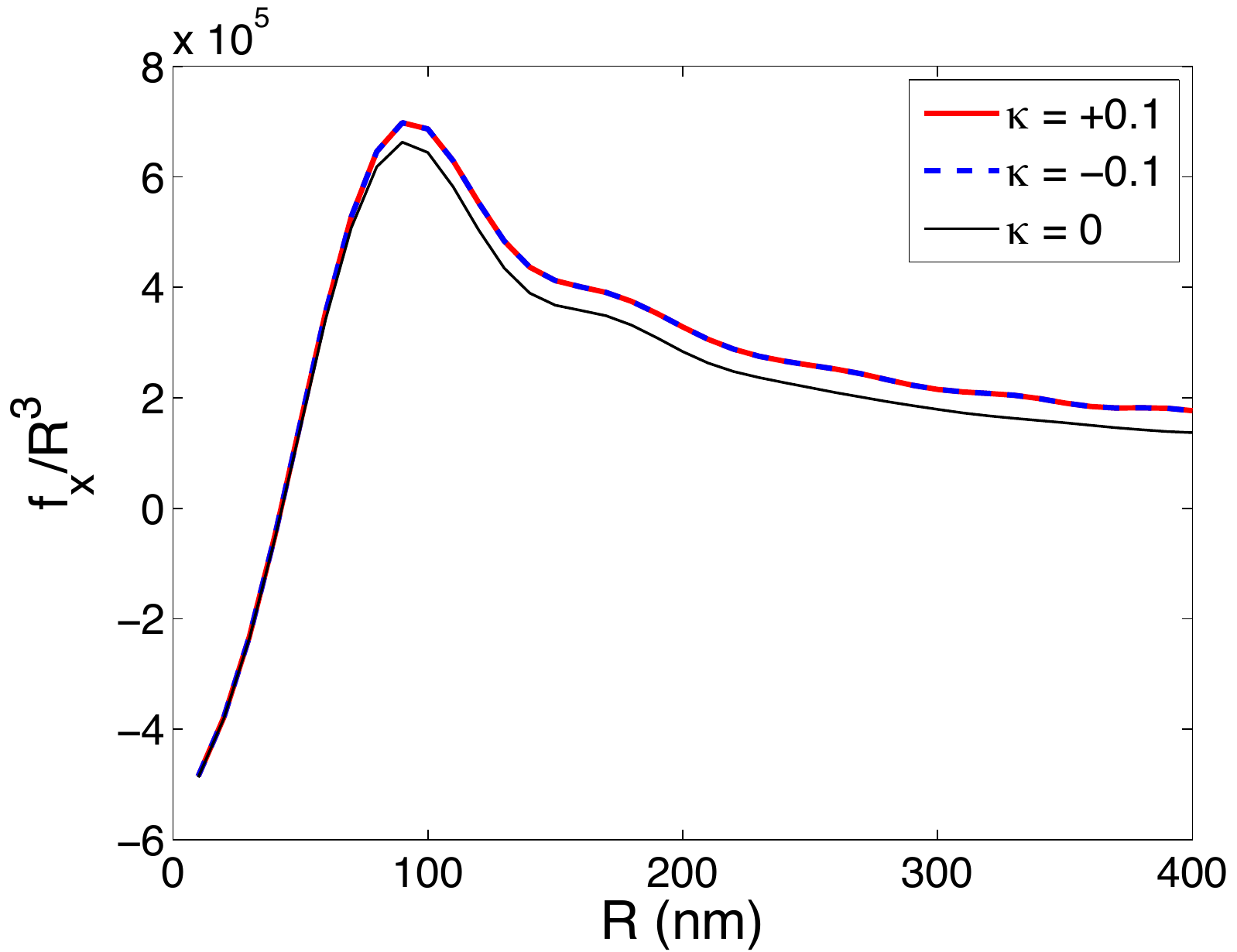}
\includegraphics[width=0.4\textwidth]{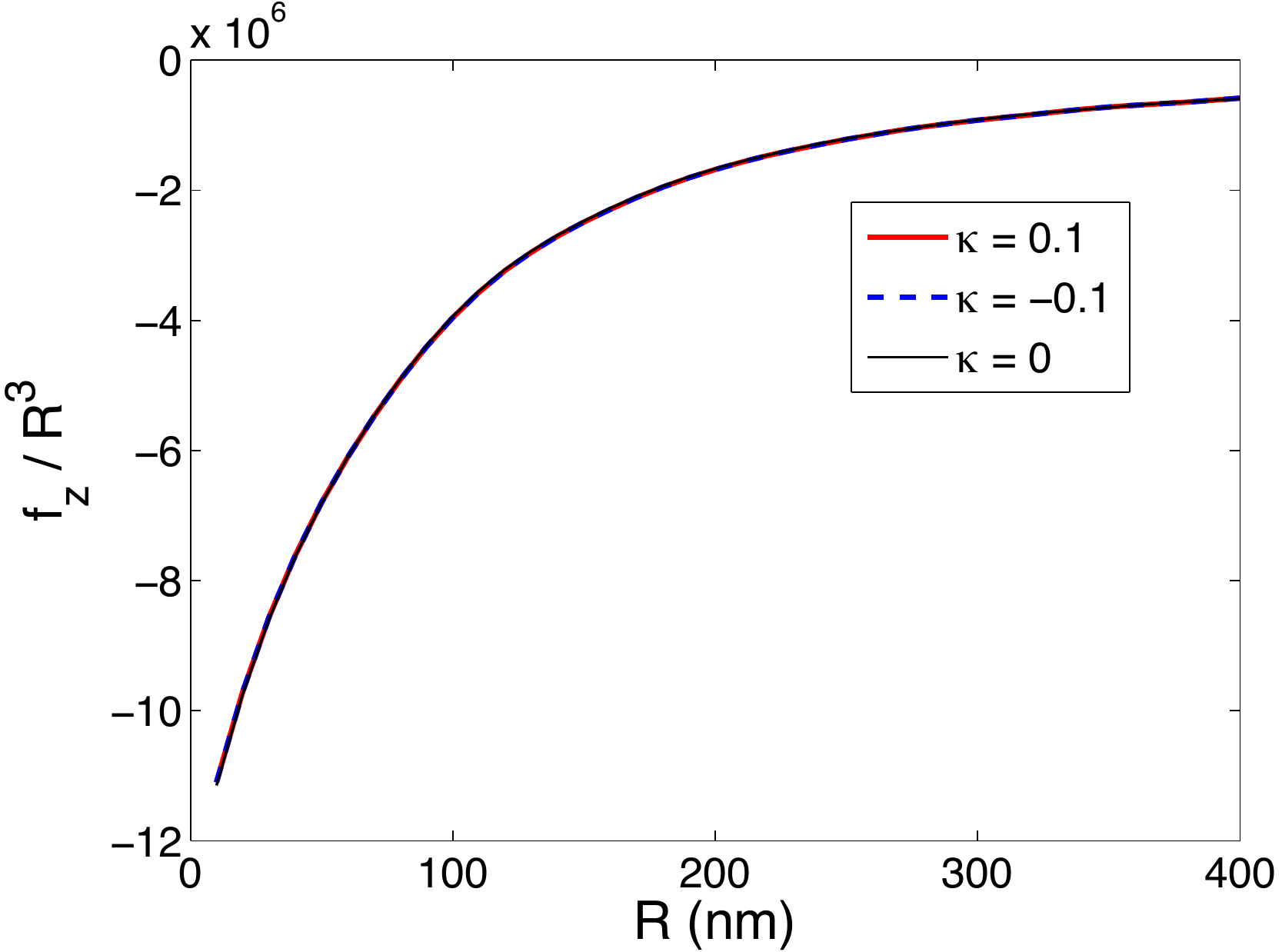}
\caption{Optical force in the longitudinal $x$ direction (top) and in the normal $z$ direction (bottom), normalized by $e^{-2q^{''}h} n_\d I_0 R^3/c$ for a non-dissipative chiral sphere ($\eps=(1.7)^2 \eps_0, \mu=\mu_0$), with optical rotation ($\kappa=\pm0.1$). The two opposite enantiomers correspond to the red (light grey) and blue (dashed line) traces, while the result for an achiral material ($\kappa=0$) is shown in black for comparison.}
 \label{fig:fxz_lmax10_dielchi}
\end{figure}

\begin{figure}[htbp]
\includegraphics[width=0.4\textwidth]{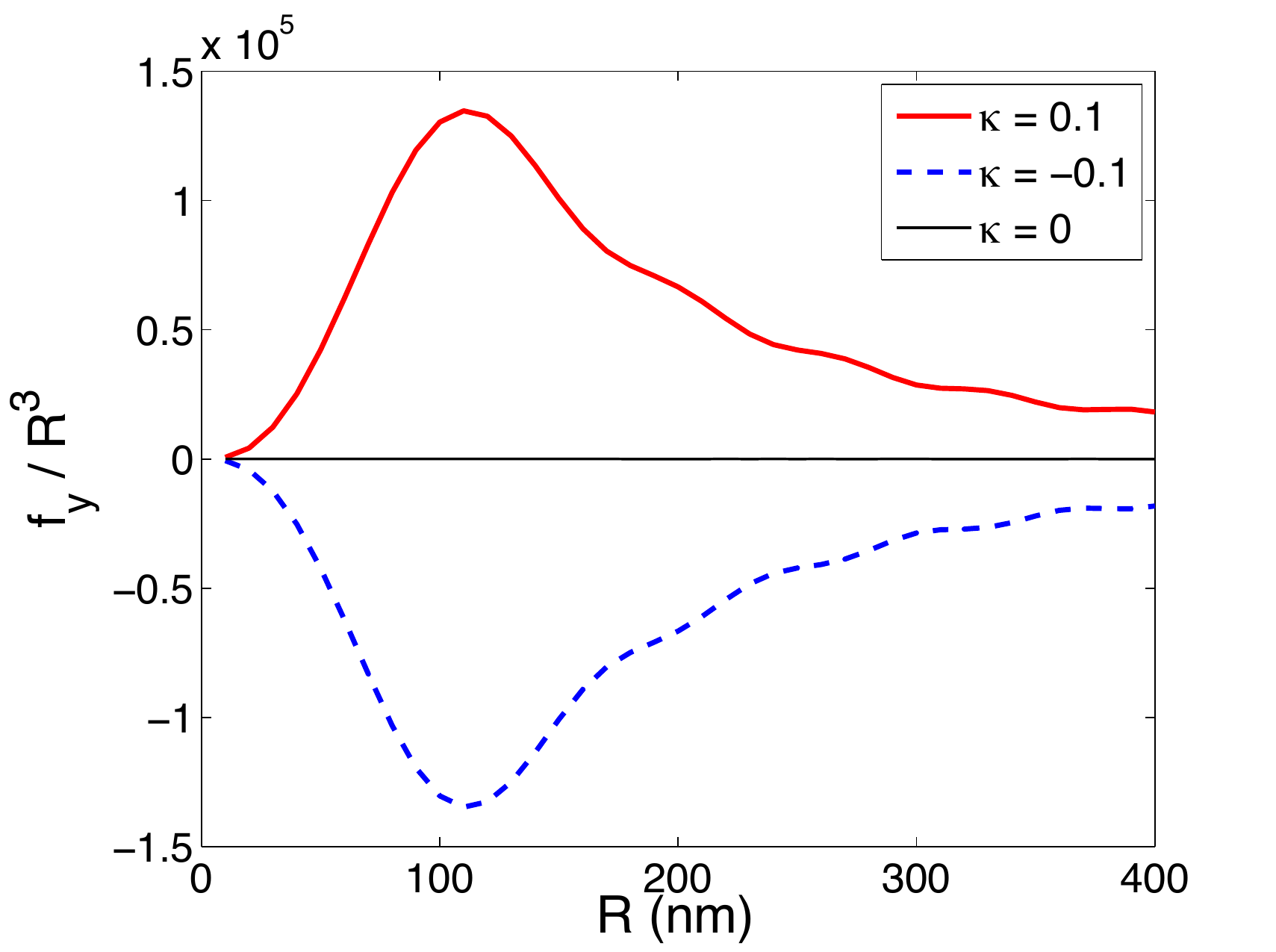}
\includegraphics[width=0.4\textwidth]{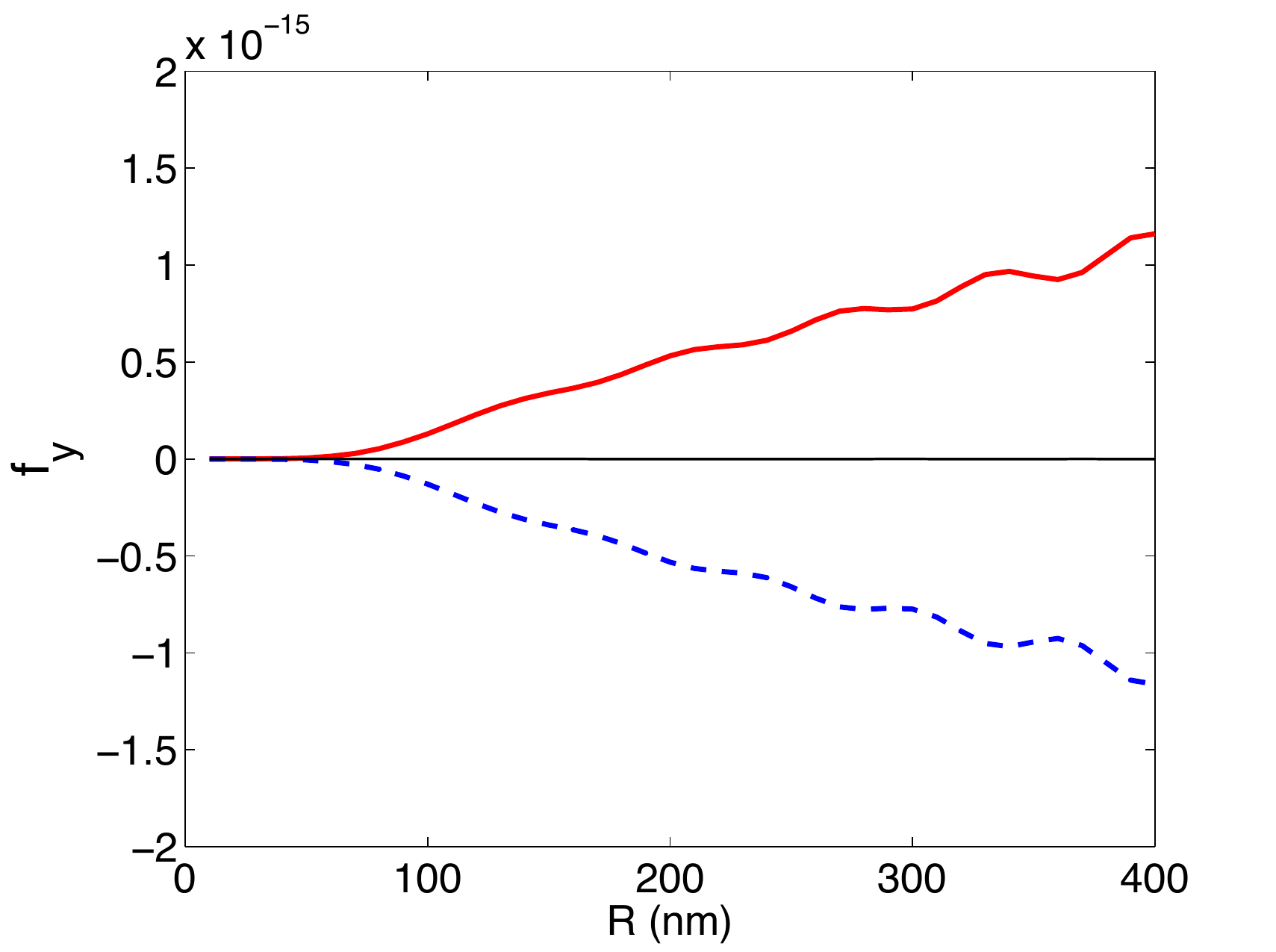}
\caption{Same as Fig.~(\ref{fig:fxz_lmax10_dielchi}) for the optical force in the transverse $y$ direction. The lower panel displays the raw term $f_y$ in order to reveal the quasi-linear dependence with respect to the sphere radius $R$. Same color code as in Fig. \ref{fig:fxz_lmax10_dielchi}. }
 \label{fig:fy_lmax10_dielchi}
\end{figure}

Let us first consider a non-dissipative dielectric chiral sphere with optical parameters $\eps=(1.7)^2 \eps_0, \mu=\mu_0$ and a real chiral parameter $\kappa=\pm0.1$ corresponding to a bulk material with optical rotation  $\left(n_+-n_-=0.2 \right)$. The results of the multipolar evaluations of the force are presented in Figs.~(\ref{fig:fxz_lmax10_dielchi},\ref{fig:fy_lmax10_dielchi}) as a function of the sphere radius. 

In the top panel of Fig.~(\ref{fig:fxz_lmax10_dielchi}),  it is interesting to note that the force is negative in the longitudinal direction for very small spheres. This comes from the fact that the sphere being non-dissipative, the radiation pressure acting on the sphere is very small and can be overtaken by the gradient force which is oriented in the $(x<0)$ direction due to the evanescence of the SP along its propagation direction. As a consequence, the particle is pulled in the backward direction due to a mere intensity gradient effect. For larger spheres, the radiation pressure increases and the force becomes positive. Multipolar effects are found in the small oscillations of the force when $R$ increases, showing the progressive onset of successive multipolar terms in the total force exerted on the sphere. We observe that the force is different for chiral and achiral spheres, but does not depend on the sign of $\kappa$. This is very reminiscent of the fact that in the dipolar polarizability $\alpha$, chirality only appears through even powers of $\kappa$. This is true for small objects, but seems to remain valid for higher multipolar terms. 

Regarding the vertical $z$ direction, which is normal to the fluid-metal interface, the force presented in the lower panel of Fig.~(\ref{fig:fxz_lmax10_dielchi}) is negative for all radii and the sphere is pulled downwards towards the interface. This is expected in the dipolar regime from the strong evanescence of the field that generates a gradient force pulling the particles downwards, but is also true for larger spheres due to the plasmonic Poynting vector (\ref{expression_PI}) that, too, is oriented downwards with $\tq' <0$. We note that the optical rotation does not seem to play an important role in the force in the vertical direction, as the results are similar for different values of $\kappa$.

 For a non-chiral object ($\kappa=0$), the force in the lateral $y$ direction is zero -see Fig.~(\ref{fig:fy_lmax10_dielchi})- as expected from the mirror symmetry of the system around the $(y=0)$-plane. However, a chiral object breaks this mirror symmetry and a lateral force appears for non-zero values of $\kappa$, whose sign also determines the direction of the force. 
The fact that the force vanishes faster than $R^3$ for small spheres shows that this force component is beyond the strict dipolar regime which only contains terms proportional to the volume of the sphere. It can be shown that the leading term is here proportional to $R^6$ as expected from the recoil force $\bF_\mathrm{recoil}$, a point that will be discussed further below. Moreover, a positive (resp. negative) value of $\kappa$ yields a positive (resp. negative) lateral force, similarly to the recoil force whose direction is determined by $\Re [\alpha \chi^*]=\Re[\alpha] \Re[\chi]$ -having here $\Im [\alpha]=0$. Interestingly for large spheres, a quasi-linearity dependence of the lateral force with respect to the sphere radius appears, as observed in the lower panel of Fig.~(\ref{fig:fy_lmax10_dielchi}).

\subsection{Forces on a dissipative dielectric sphere with circular dichroism}

\begin{figure}[htbp]
\includegraphics[width=0.4\textwidth]{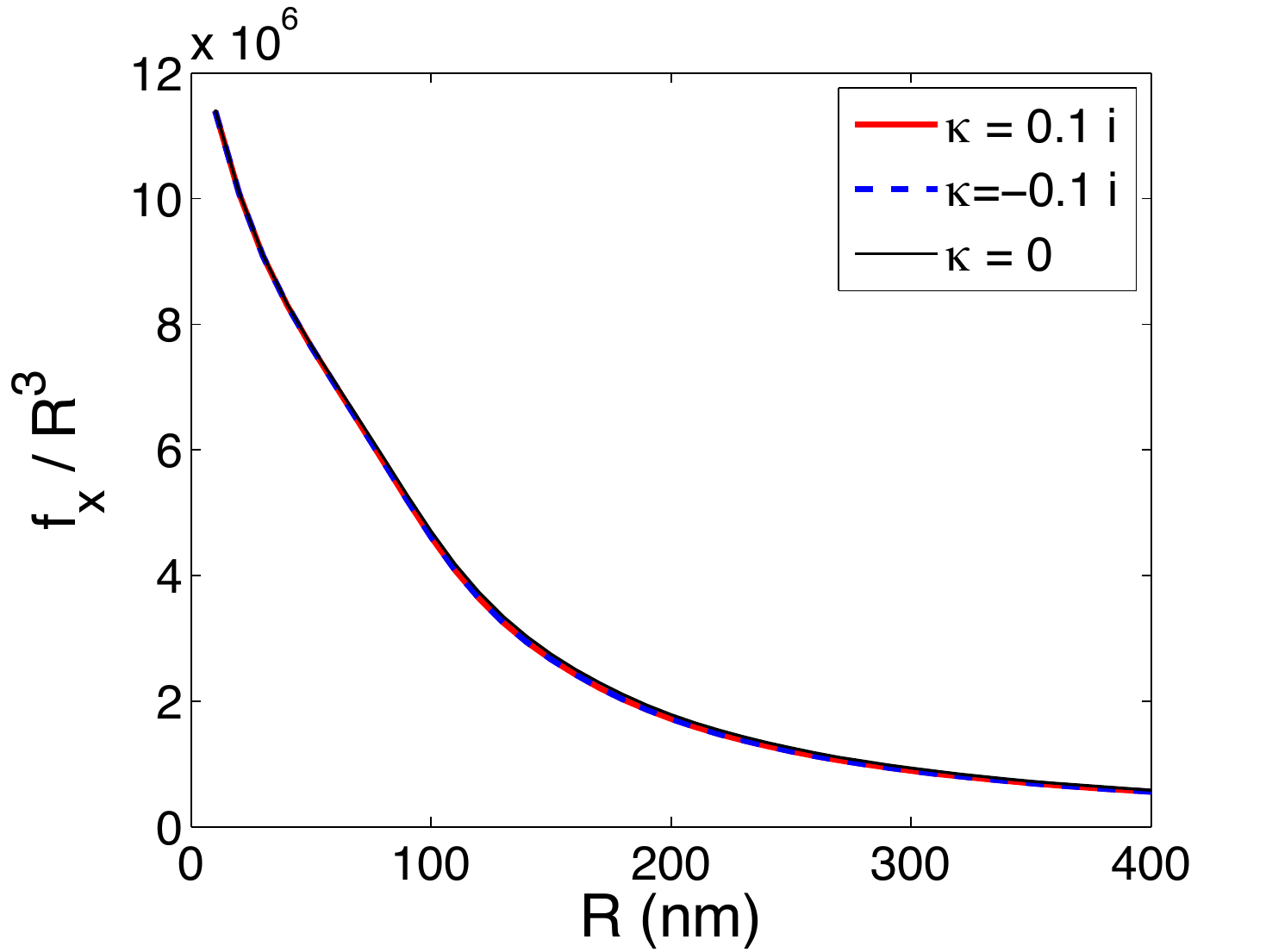}
\includegraphics[width=0.4\textwidth]{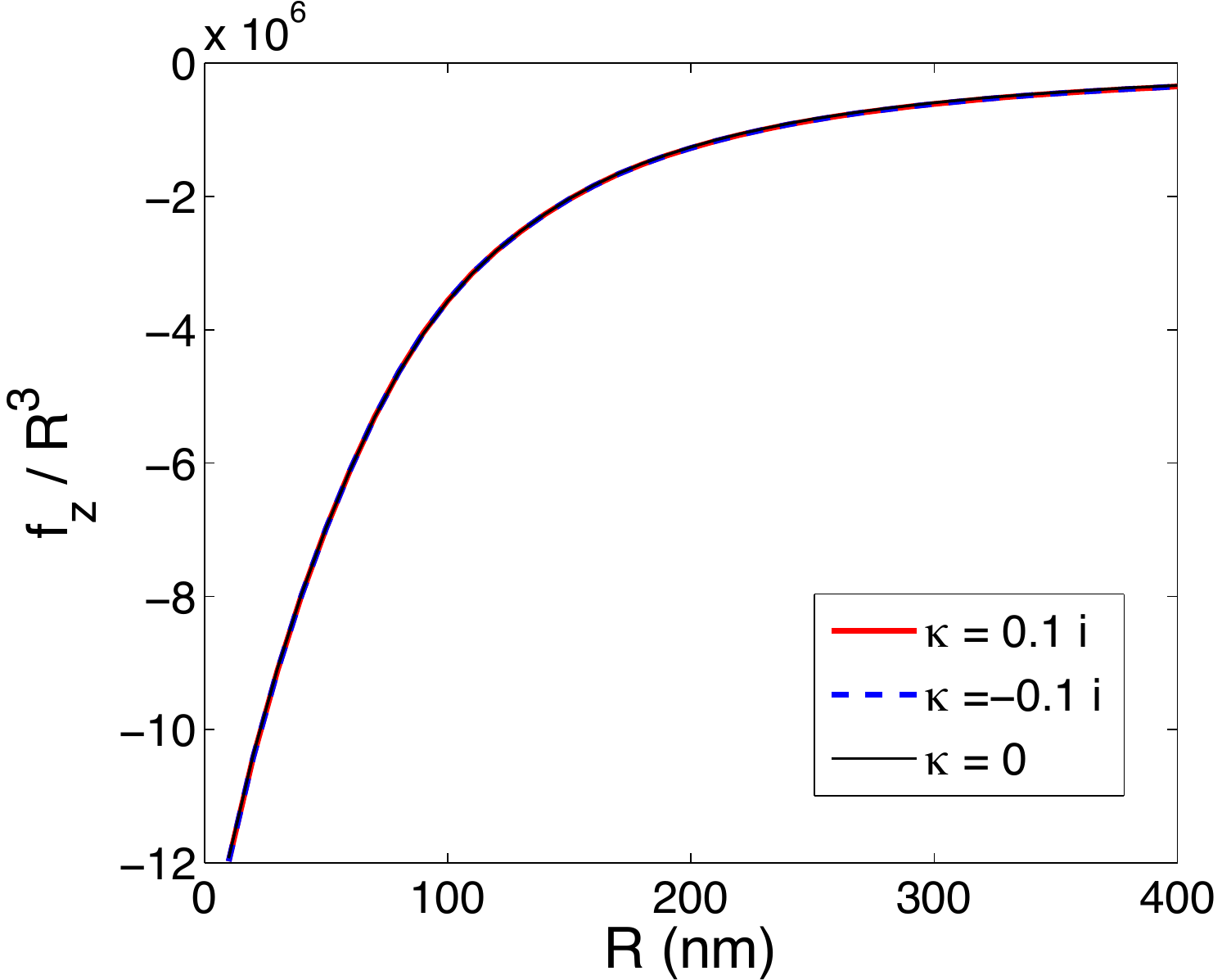}
\caption{Optical force in the longitudinal $x$ direction (top) and normal $z$ direction (bottom), normalized by $e^{-2q^{''}h} n_\d I_0 R^3/c$ for a dissipative chiral sphere ($\eps=(1.7+0.1\imath)^2 \eps_0, \mu=\mu_0$), with circular dichroism ($\kappa=\pm 0.1\imath$). The two opposite enantiomers correspond to red (light grey) and blue (dashed line) traces, while the result for an achiral material ($\kappa=0$) is shown in black for comparison.}
 \label{fig:fxz_lmax10_dielchi2}
\end{figure}

\begin{figure}[htbp]
\includegraphics[width=0.4\textwidth]{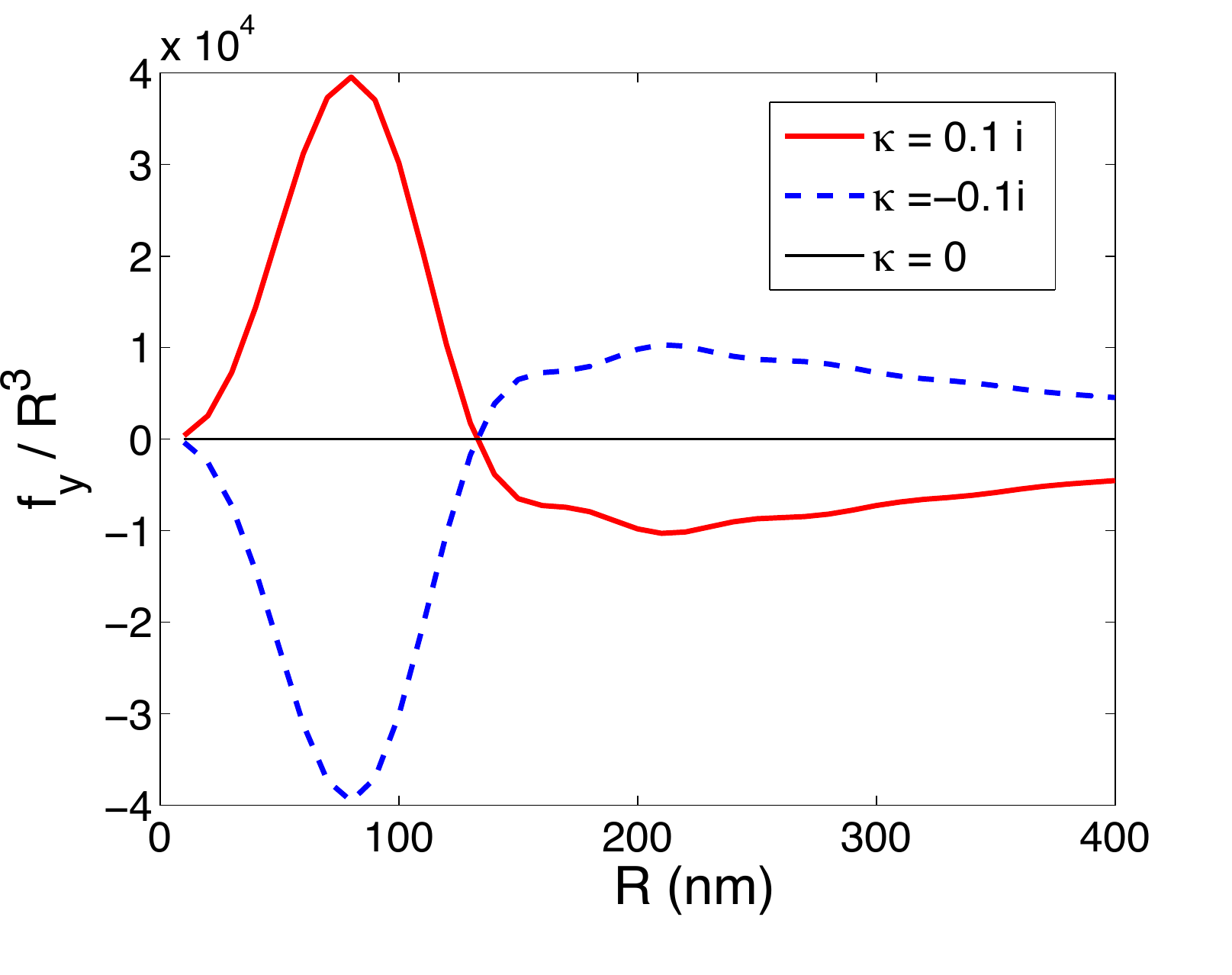}
\caption{Same as Fig.~(\ref{fig:fxz_lmax10_dielchi2}) for the optical force in the transverse $y$ direction. Same color code as in Fig. \ref{fig:fxz_lmax10_dielchi2}.}
 \label{fig:fy_lmax10_dielchi2}
\end{figure}

We now consider dissipative dielectric spheres ($\eps=(1.7+0.1\imath)^2 \eps_0, \mu=\mu_0$) with imaginary chiral parameter $\kappa=\pm0.1\imath$ corresponding to a bulk material with circular dichroism  $\left(n_+-n_-=0.2 \imath \right)$. The results of the multipolar evaluation of the force are presented in Figs.~(\ref{fig:fxz_lmax10_dielchi2},\ref{fig:fy_lmax10_dielchi2}) as a function of the sphere radius. 
For the $x$ and $z$ direction, circular dichroism does not have any noticeable effect on the force, as seen in both panels of Fig.~(\ref{fig:fxz_lmax10_dielchi2}), as the curves superimpose regardless of the value of $\kappa$. However, for the lateral force presented in Fig.~(\ref{fig:fy_lmax10_dielchi2}) we observe two non-trivial effects of chirality: first, the force $f_y$ is again non-zero and points in opposite directions for opposite enantiomers, and then, the lateral force flips its sign for increasing values of the sphere radius.  
The first observation corroborates the results obtained with a model dipole from the recoil force (\ref{F_recoil}), which predicts the appearance of a lateral force whose direction is given by the sign of $\Re[\alpha \chi^*] \simeq \Im[\alpha] \Im[\chi]$. For small spheres ($R \lesssim 100$ nm), we then recover the prediction of a positive (resp. negative) lateral force for positive (resp. negative) values of $\Im[\kappa]$. Nevertheless, the sign flip shows that the result can be opposite for larger spheres. This sign flip results from the interplay between multipolar components and the dynamical coupling with chirality, as it has already been demonstrated for propagative wave in configurations where chiral momentum transfers were the only driving force \cite{canaguier2015chiral}.

\subsection{Comparison of the recoil term with multipolar calculations}

As the last point of the article, we assess the accuracy of the prediction provided by the recoil optical force -Eq. (\ref{F_recoil_pm})- in the situation where the dipole represents a nanosphere in the small-size limit. To do so, we confront this recoil term to a multipolar evaluation of the optical force. For all different sets of conditions, we first checked that the force $\bF$ derived from a multipolar calculation and the force $\bF_\dip$ give the same result up to order $R^3$. This further confirms the validity of this strict dipolar approach as the first order expansion of the force acting on a nanoparticle in the small-size limit. We then investigate next-to-leading order (NTLO) terms by taking the difference $(\bF-\bF_\dip)$ between the multipolar result and the strict dipolar term, and comparing it to the corrective recoil force $\bF_\mathrm{recoil}$. Both evaluations of the forces are done for the case of Figs.~(\ref{fig:fxz_lmax10_dielchi},\ref{fig:fy_lmax10_dielchi}) of a non-dissipative dielectric sphere with optical rotation. To analyze the physical nature of the beyond-dipolar terms at play, we introduce the quantity
\begin{align*}
a_R = \frac{\d \ln f}{\d \ln R}
\end{align*}
that measures the local power law in $R$ for the force, fitting locally the curvature of $f(R)$ with a power function $R^{a_R}$.

\begin{figure}[htbp]
\includegraphics[width=0.45\textwidth]{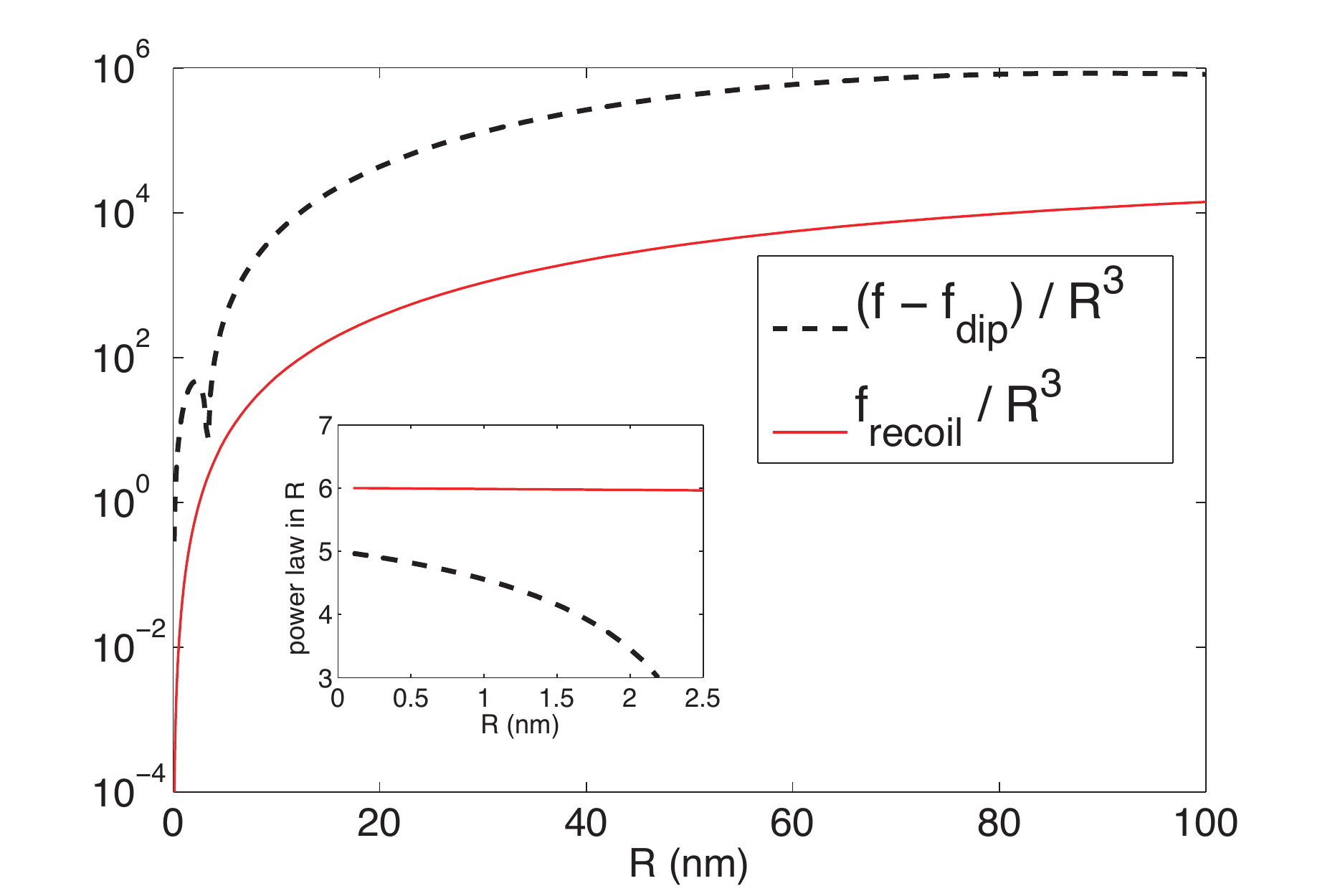}
\includegraphics[width=0.45\textwidth]{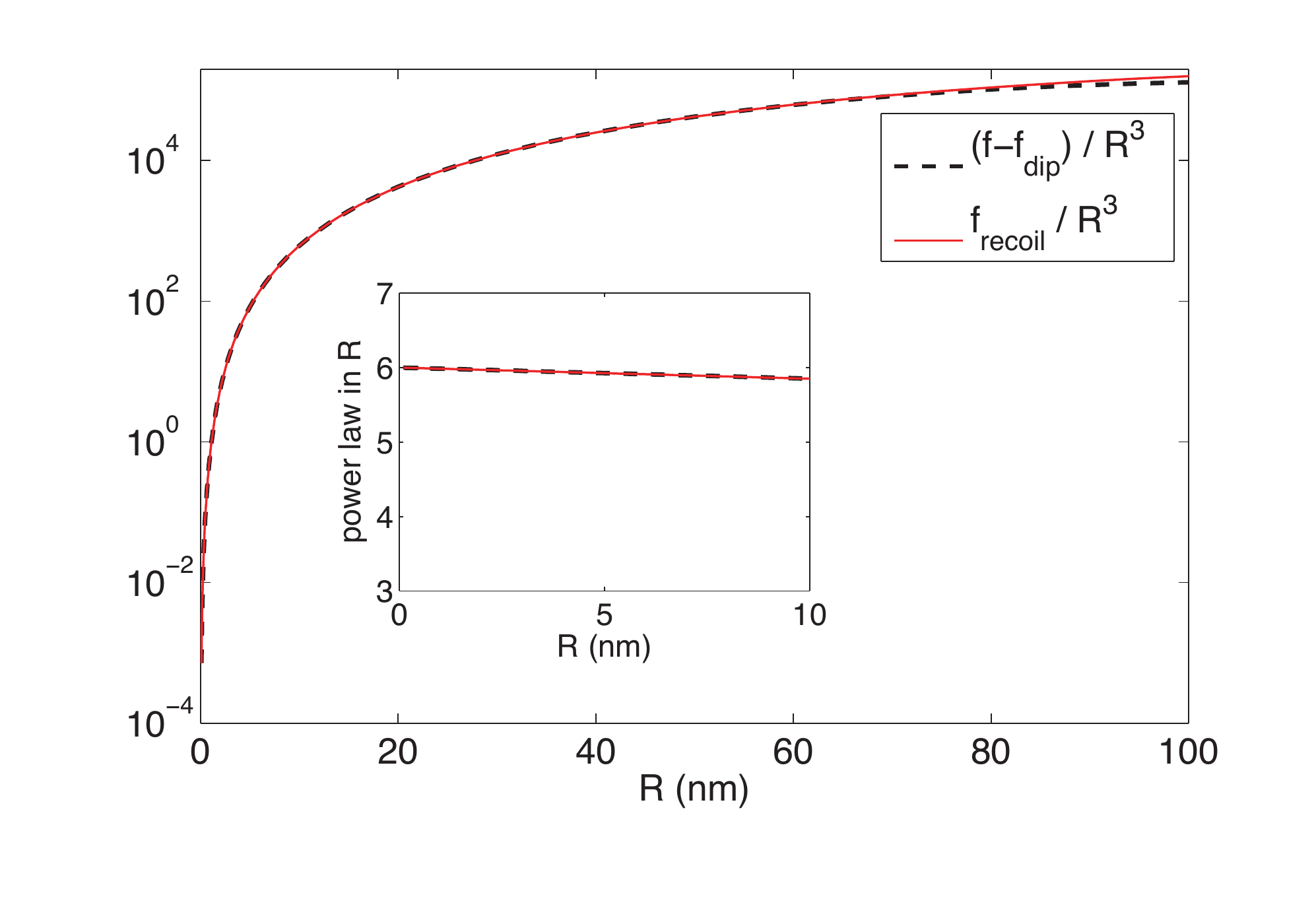}
\caption{Comparison of the NTLO results $\bF-\bF_\dip$ and $\bF_\mathrm{recoil}$ for the optical force in the $x$ direction (top) and in the lateral $y$ direction (bottom). The insets shows the limit of the local power law in $R$ in the small-size limit, which give information on the nature of the NTLO term. The sphere is made of a non-dissipative dielectric ($\eps=(1.7)^2 \eps_0, \mu=\mu_0$), with optical rotation ($\kappa=\pm0.1$).}
 \label{fig:f_NTLO_dielchi}
\end{figure}

First we present in the top panel of Fig.~(\ref{fig:f_NTLO_dielchi}) the results obtained for the force in the $x$ direction. 
We find that the two quantities $(\bF-\bF_\dip)$ and $\bF_\mathrm{recoil}$  differ, even in the limit of small spheres. This is due to the fact that the NTLO term scales as $R^5$ for the multipolar result (as seen in the inset) coming from electric and magnetic quadrupolar terms, whereas the recoil force only gives a contribution scaling with $R^6$ coming from self-interaction of dipolar moments. A similar observation can be made regarding the force in the $z$ direction. This difference shows that adding the recoil force to the dipolar result $\bF_\dip$ does not bring additional information or better accuracy in the evaluation of the force in the small sphere limit, because it misses quadrupolar contributions that are dominant over the recoil terms. The same results are obtained when considering the case of a dissipative dielectric nanosphere with circular dichroism as in Figs.~(\ref{fig:fxz_lmax10_dielchi2},\ref{fig:fy_lmax10_dielchi2}).

We then look at the force in the $y$ direction in the lower panel of Fig.~(\ref{fig:f_NTLO_dielchi}). In this case, the strictly dipolar result gives a null contribution, whereas the recoil force predicts a force scaling as $R^6$. The multipolar result agrees with the recoil force, with the same term in $R^6$ as leading contribution coming from the self interaction of dipole moments. This gives the final confirmation on the existence of such lateral chiral forces on a nanosphere, in agreement with the recoil term derived within a dipole model. This agreement between the two quantities comes from the fact that electric and magnetic quadripolar moments do not have any direct contribution to the force in the lateral direction. Otherwise, they would necessarily impact the contribution of the recoil terms with respect to the whole expression of the force. We emphasize that this point  could not be taken as an \textit{a priori} assumption. By looking at the quantity $\bF-\bF_\dip-\bF_\mathrm{recoil}$ we see that the next term in the small-size expansion after the recoil force is proportional to $R^8$, thus coming from the self interaction between dipolar and quadrupolar moments. This shows that quadrupolar moments do have an influence on the lateral force, but only indirectly through their interaction with dipolar moments. Again, we checked that all these conclusions still hold when taking a dissipative dielectric nanosphere with circular dichroism.  

\section{Conclusion}

We have shown that a plasmonic field propagating on a metal-dielectric interface is able to induce a lateral force on chiral spheres with either optical rotation or circular dichroism. This lateral force points in opposite directions for two enantiomeres of the same size and opens the possibility to use plasmonic fields for optical enantioseparation. Physically, the lateral force comes from the transfer of plasmonic transverse spin angular momentum into transverse linear momentum for the particle, transfer being mediated by the chirality of the dipole. 

Our multipolar analysis shows that as soon as one is interested in real finite size objects, the inclusion of a recoil term in the dipolar force does not always lead to an improved accuracy in the evaluation of the force, as quadrupolar terms must be included in the first place in the beyond-dipolar expansion. Moreover, we observe for larger sphere a sign inversion of the lateral force which cannot be seen when merely considering a chiral dipole. This reversal is reminiscent of multipolar effects on angular-to-linear cross momentum transfers mediated by chirality that we have already discussed in the context of pulling forces or left-handed torques \cite{canaguier2015chiral}. Such sign inversions could have important consequences when designing discriminatory protocols based on optical chiral forces.

\section{Acknowledgments}

We acknowledge support from the French program Investissement d’Avenir (ANR, Equipex Union).

\bibliography{biblio_lateral_plasmonic}

\begin{thebibliography}{32}%
\makeatletter
\providecommand \@ifxundefined [1]{%
 \@ifx{#1\undefined}
}%
\providecommand \@ifnum [1]{%
 \ifnum #1\expandafter \@firstoftwo
 \else \expandafter \@secondoftwo
 \fi
}%
\providecommand \@ifx [1]{%
 \ifx #1\expandafter \@firstoftwo
 \else \expandafter \@secondoftwo
 \fi
}%
\providecommand \natexlab [1]{#1}%
\providecommand \enquote  [1]{``#1''}%
\providecommand \bibnamefont  [1]{#1}%
\providecommand \bibfnamefont [1]{#1}%
\providecommand \citenamefont [1]{#1}%
\providecommand \href@noop [0]{\@secondoftwo}%
\providecommand \href [0]{\begingroup \@sanitize@url \@href}%
\providecommand \@href[1]{\@@startlink{#1}\@@href}%
\providecommand \@@href[1]{\endgroup#1\@@endlink}%
\providecommand \@sanitize@url [0]{\catcode `\\12\catcode `\$12\catcode
  `\&12\catcode `\#12\catcode `\^12\catcode `\_12\catcode `\%12\relax}%
\providecommand \@@startlink[1]{}%
\providecommand \@@endlink[0]{}%
\providecommand \url  [0]{\begingroup\@sanitize@url \@url }%
\providecommand \@url [1]{\endgroup\@href {#1}{\urlprefix }}%
\providecommand \urlprefix  [0]{URL }%
\providecommand \Eprint [0]{\href }%
\providecommand \doibase [0]{http://dx.doi.org/}%
\providecommand \selectlanguage [0]{\@gobble}%
\providecommand \bibinfo  [0]{\@secondoftwo}%
\providecommand \bibfield  [0]{\@secondoftwo}%
\providecommand \translation [1]{[#1]}%
\providecommand \BibitemOpen [0]{}%
\providecommand \bibitemStop [0]{}%
\providecommand \bibitemNoStop [0]{.\EOS\space}%
\providecommand \EOS [0]{\spacefactor3000\relax}%
\providecommand \BibitemShut  [1]{\csname bibitem#1\endcsname}%
\let\auto@bib@innerbib\@empty
\bibitem [{\citenamefont {Canaguier-Durand}\ \emph
  {et~al.}(2013{\natexlab{a}})\citenamefont {Canaguier-Durand}, \citenamefont
  {Hutchison}, \citenamefont {Genet},\ and\ \citenamefont
  {Ebbesen}}]{canaguier2013mechanical}%
  \BibitemOpen
  \bibfield  {author} {\bibinfo {author} {\bibfnamefont {A.}~\bibnamefont
  {Canaguier-Durand}}, \bibinfo {author} {\bibfnamefont {J.~A.}\ \bibnamefont
  {Hutchison}}, \bibinfo {author} {\bibfnamefont {C.}~\bibnamefont {Genet}}, \
  and\ \bibinfo {author} {\bibfnamefont {T.~W.}\ \bibnamefont {Ebbesen}},\
  }\href@noop {} {\bibfield  {journal} {\bibinfo  {journal} {New J. Phys.}\
  }\textbf {\bibinfo {volume} {15}},\ \bibinfo {pages} {123037} (\bibinfo
  {year} {2013}{\natexlab{a}})}\BibitemShut {NoStop}%
\bibitem [{\citenamefont {Cameron}\ \emph
  {et~al.}(2014{\natexlab{a}})\citenamefont {Cameron}, \citenamefont
  {Barnett},\ and\ \citenamefont {Yao}}]{cameron2014discriminatory}%
  \BibitemOpen
  \bibfield  {author} {\bibinfo {author} {\bibfnamefont {R.~P.}\ \bibnamefont
  {Cameron}}, \bibinfo {author} {\bibfnamefont {S.~M.}\ \bibnamefont
  {Barnett}}, \ and\ \bibinfo {author} {\bibfnamefont {A.~M.}\ \bibnamefont
  {Yao}},\ }\href@noop {} {\bibfield  {journal} {\bibinfo  {journal} {New J.
  Phys.}\ }\textbf {\bibinfo {volume} {16}},\ \bibinfo {pages} {013020}
  (\bibinfo {year} {2014}{\natexlab{a}})}\BibitemShut {NoStop}%
\bibitem [{\citenamefont {Cameron}\ \emph
  {et~al.}(2014{\natexlab{b}})\citenamefont {Cameron}, \citenamefont {Yao},\
  and\ \citenamefont {Barnett}}]{cameron2014diffraction}%
  \BibitemOpen
  \bibfield  {author} {\bibinfo {author} {\bibfnamefont {R.~P.}\ \bibnamefont
  {Cameron}}, \bibinfo {author} {\bibfnamefont {A.~M.}\ \bibnamefont {Yao}}, \
  and\ \bibinfo {author} {\bibfnamefont {S.~M.}\ \bibnamefont {Barnett}},\
  }\href@noop {} {\bibfield  {journal} {\bibinfo  {journal} {J. Phys. Chem. A}\
  }\textbf {\bibinfo {volume} {118}},\ \bibinfo {pages} {3472} (\bibinfo {year}
  {2014}{\natexlab{b}})}\BibitemShut {NoStop}%
\bibitem [{\citenamefont {Bradshaw}\ and\ \citenamefont
  {Andrews}(2014)}]{bradshaw2014chiral}%
  \BibitemOpen
  \bibfield  {author} {\bibinfo {author} {\bibfnamefont {D.~S.}\ \bibnamefont
  {Bradshaw}}\ and\ \bibinfo {author} {\bibfnamefont {D.~L.}\ \bibnamefont
  {Andrews}},\ }\href@noop {} {\bibfield  {journal} {\bibinfo  {journal} {New
  Journal of Physics}\ }\textbf {\bibinfo {volume} {16}},\ \bibinfo {pages}
  {103021} (\bibinfo {year} {2014})}\BibitemShut {NoStop}%
\bibitem [{\citenamefont {Tkachenko}\ and\ \citenamefont
  {Brasselet}(2013)}]{tkachenko2013spin}%
  \BibitemOpen
  \bibfield  {author} {\bibinfo {author} {\bibfnamefont {G.}~\bibnamefont
  {Tkachenko}}\ and\ \bibinfo {author} {\bibfnamefont {E.}~\bibnamefont
  {Brasselet}},\ }\href@noop {} {\bibfield  {journal} {\bibinfo  {journal}
  {Phys. Rev. Lett.}\ }\textbf {\bibinfo {volume} {111}},\ \bibinfo {pages}
  {033605} (\bibinfo {year} {2013})}\BibitemShut {NoStop}%
\bibitem [{\citenamefont {Tkachenko}\ and\ \citenamefont
  {Brasselet}(2014{\natexlab{a}})}]{tkachenko2014optofluidic}%
  \BibitemOpen
  \bibfield  {author} {\bibinfo {author} {\bibfnamefont {G.}~\bibnamefont
  {Tkachenko}}\ and\ \bibinfo {author} {\bibfnamefont {E.}~\bibnamefont
  {Brasselet}},\ }\href@noop {} {\bibfield  {journal} {\bibinfo  {journal}
  {Nat. Commun.}\ }\textbf {\bibinfo {volume} {5}},\ \bibinfo {pages} {3577}
  (\bibinfo {year} {2014}{\natexlab{a}})}\BibitemShut {NoStop}%
\bibitem [{\citenamefont {Tkachenko}\ and\ \citenamefont
  {Brasselet}(2014{\natexlab{b}})}]{tkachenko2014helicity}%
  \BibitemOpen
  \bibfield  {author} {\bibinfo {author} {\bibfnamefont {G.}~\bibnamefont
  {Tkachenko}}\ and\ \bibinfo {author} {\bibfnamefont {E.}~\bibnamefont
  {Brasselet}},\ }\href@noop {} {\bibfield  {journal} {\bibinfo  {journal}
  {Nat. Commun.}\ }\textbf {\bibinfo {volume} {5}},\ \bibinfo {pages} {4491}
  (\bibinfo {year} {2014}{\natexlab{b}})}\BibitemShut {NoStop}%
\bibitem [{\citenamefont {Ding}\ \emph {et~al.}(2014)\citenamefont {Ding},
  \citenamefont {Ng}, \citenamefont {Zhou},\ and\ \citenamefont
  {Chan}}]{ding2014realization}%
  \BibitemOpen
  \bibfield  {author} {\bibinfo {author} {\bibfnamefont {K.}~\bibnamefont
  {Ding}}, \bibinfo {author} {\bibfnamefont {J.}~\bibnamefont {Ng}}, \bibinfo
  {author} {\bibfnamefont {L.}~\bibnamefont {Zhou}}, \ and\ \bibinfo {author}
  {\bibfnamefont {C.}~\bibnamefont {Chan}},\ }\href@noop {} {\bibfield
  {journal} {\bibinfo  {journal} {Phys. Rev. A}\ }\textbf {\bibinfo {volume}
  {89}},\ \bibinfo {pages} {063825} (\bibinfo {year} {2014})}\BibitemShut
  {NoStop}%
\bibitem [{\citenamefont {Wang}\ and\ \citenamefont
  {Chan}(2014)}]{wang2014lateral}%
  \BibitemOpen
  \bibfield  {author} {\bibinfo {author} {\bibfnamefont {S.}~\bibnamefont
  {Wang}}\ and\ \bibinfo {author} {\bibfnamefont {C.}~\bibnamefont {Chan}},\
  }\href@noop {} {\bibfield  {journal} {\bibinfo  {journal} {Nat. Commun.}\
  }\textbf {\bibinfo {volume} {5}},\ \bibinfo {pages} {3307} (\bibinfo {year}
  {2014})}\BibitemShut {NoStop}%
\bibitem [{\citenamefont {Chen}\ \emph {et~al.}(2014)\citenamefont {Chen},
  \citenamefont {Wang}, \citenamefont {Lu}, \citenamefont {Liu},\ and\
  \citenamefont {Lin}}]{chen2014tailoring}%
  \BibitemOpen
  \bibfield  {author} {\bibinfo {author} {\bibfnamefont {H.}~\bibnamefont
  {Chen}}, \bibinfo {author} {\bibfnamefont {N.}~\bibnamefont {Wang}}, \bibinfo
  {author} {\bibfnamefont {W.}~\bibnamefont {Lu}}, \bibinfo {author}
  {\bibfnamefont {S.}~\bibnamefont {Liu}}, \ and\ \bibinfo {author}
  {\bibfnamefont {Z.}~\bibnamefont {Lin}},\ }\href@noop {} {\bibfield
  {journal} {\bibinfo  {journal} {Phys. Rev. A}\ }\textbf {\bibinfo {volume}
  {90}},\ \bibinfo {pages} {043850} (\bibinfo {year} {2014})}\BibitemShut
  {NoStop}%
\bibitem [{\citenamefont {Hayat}\ \emph {et~al.}(2014)\citenamefont {Hayat},
  \citenamefont {M{\"u}ller},\ and\ \citenamefont
  {Capasso}}]{hayat2014lateral}%
  \BibitemOpen
  \bibfield  {author} {\bibinfo {author} {\bibfnamefont {A.}~\bibnamefont
  {Hayat}}, \bibinfo {author} {\bibfnamefont {J.}~\bibnamefont {M{\"u}ller}}, \
  and\ \bibinfo {author} {\bibfnamefont {F.}~\bibnamefont {Capasso}},\
  }\href@noop {} {\bibfield  {journal} {\bibinfo  {journal} {arXiv preprint
  arXiv:1408.2268}\ } (\bibinfo {year} {2014})}\BibitemShut {NoStop}%
\bibitem [{\citenamefont {Canaguier-Durand}\ and\ \citenamefont
  {Genet}(2014{\natexlab{a}})}]{canaguier2014chiral}%
  \BibitemOpen
  \bibfield  {author} {\bibinfo {author} {\bibfnamefont {A.}~\bibnamefont
  {Canaguier-Durand}}\ and\ \bibinfo {author} {\bibfnamefont {C.}~\bibnamefont
  {Genet}},\ }\href@noop {} {\bibfield  {journal} {\bibinfo  {journal} {Phys.
  Rev. A}\ }\textbf {\bibinfo {volume} {90}},\ \bibinfo {pages} {023842}
  (\bibinfo {year} {2014}{\natexlab{a}})}\BibitemShut {NoStop}%
\bibitem [{\citenamefont {Alizadeh}\ and\ \citenamefont
  {Reinhard}(2015)}]{alizadeh2015plasmonically}%
  \BibitemOpen
  \bibfield  {author} {\bibinfo {author} {\bibfnamefont {M.}~\bibnamefont
  {Alizadeh}}\ and\ \bibinfo {author} {\bibfnamefont {B.~M.}\ \bibnamefont
  {Reinhard}},\ }\href@noop {} {\bibfield  {journal} {\bibinfo  {journal} {ACS
  Photonics}\ }\textbf {\bibinfo {volume} {2}},\ \bibinfo {pages} {361}
  (\bibinfo {year} {2015})}\BibitemShut {NoStop}%
\bibitem [{\citenamefont {Canaguier-Durand}\ and\ \citenamefont
  {Genet}(2015)}]{canaguier2015chiral}%
  \BibitemOpen
  \bibfield  {author} {\bibinfo {author} {\bibfnamefont {A.}~\bibnamefont
  {Canaguier-Durand}}\ and\ \bibinfo {author} {\bibfnamefont {C.}~\bibnamefont
  {Genet}},\ }\href@noop {} {\bibfield  {journal} {\bibinfo  {journal} {arXiv
  preprint arXiv:1503.02175}\ } (\bibinfo {year} {2015})}\BibitemShut {NoStop}%
\bibitem [{\citenamefont {Yan}\ \emph {et~al.}(2012)\citenamefont {Yan},
  \citenamefont {Xu}, \citenamefont {Xu}, \citenamefont {Ma}, \citenamefont
  {Kuang}, \citenamefont {Wang},\ and\ \citenamefont {Kotov}}]{yan2012self}%
  \BibitemOpen
  \bibfield  {author} {\bibinfo {author} {\bibfnamefont {W.}~\bibnamefont
  {Yan}}, \bibinfo {author} {\bibfnamefont {L.}~\bibnamefont {Xu}}, \bibinfo
  {author} {\bibfnamefont {C.}~\bibnamefont {Xu}}, \bibinfo {author}
  {\bibfnamefont {W.}~\bibnamefont {Ma}}, \bibinfo {author} {\bibfnamefont
  {H.}~\bibnamefont {Kuang}}, \bibinfo {author} {\bibfnamefont
  {L.}~\bibnamefont {Wang}}, \ and\ \bibinfo {author} {\bibfnamefont {N.~A.}\
  \bibnamefont {Kotov}},\ }\href@noop {} {\bibfield  {journal} {\bibinfo
  {journal} {Journal of the American Chemical Society}\ }\textbf {\bibinfo
  {volume} {134}},\ \bibinfo {pages} {15114} (\bibinfo {year}
  {2012})}\BibitemShut {NoStop}%
\bibitem [{\citenamefont {McPeak}\ \emph {et~al.}(2014)\citenamefont {McPeak},
  \citenamefont {van Engers}, \citenamefont {Blome}, \citenamefont {Park},
  \citenamefont {Burger}, \citenamefont {Gosalvez}, \citenamefont {Faridi},
  \citenamefont {Ries}, \citenamefont {Sahu},\ and\ \citenamefont
  {Norris}}]{mcpeak2014complex}%
  \BibitemOpen
  \bibfield  {author} {\bibinfo {author} {\bibfnamefont {K.~M.}\ \bibnamefont
  {McPeak}}, \bibinfo {author} {\bibfnamefont {C.~D.}\ \bibnamefont {van
  Engers}}, \bibinfo {author} {\bibfnamefont {M.}~\bibnamefont {Blome}},
  \bibinfo {author} {\bibfnamefont {J.~H.}\ \bibnamefont {Park}}, \bibinfo
  {author} {\bibfnamefont {S.}~\bibnamefont {Burger}}, \bibinfo {author}
  {\bibfnamefont {M.~A.}\ \bibnamefont {Gosalvez}}, \bibinfo {author}
  {\bibfnamefont {A.}~\bibnamefont {Faridi}}, \bibinfo {author} {\bibfnamefont
  {Y.~R.}\ \bibnamefont {Ries}}, \bibinfo {author} {\bibfnamefont
  {A.}~\bibnamefont {Sahu}}, \ and\ \bibinfo {author} {\bibfnamefont {D.~J.}\
  \bibnamefont {Norris}},\ }\href@noop {} {\bibfield  {journal} {\bibinfo
  {journal} {Nano letters}\ }\textbf {\bibinfo {volume} {14}},\ \bibinfo
  {pages} {2934} (\bibinfo {year} {2014})}\BibitemShut {NoStop}%
\bibitem [{\citenamefont {Bliokh}\ \emph {et~al.}(2014)\citenamefont {Bliokh},
  \citenamefont {Bekshaev},\ and\ \citenamefont
  {Nori}}]{bliokh2014extraordinary}%
  \BibitemOpen
  \bibfield  {author} {\bibinfo {author} {\bibfnamefont {K.~Y.}\ \bibnamefont
  {Bliokh}}, \bibinfo {author} {\bibfnamefont {A.~Y.}\ \bibnamefont
  {Bekshaev}}, \ and\ \bibinfo {author} {\bibfnamefont {F.}~\bibnamefont
  {Nori}},\ }\href@noop {} {\bibfield  {journal} {\bibinfo  {journal} {Nat.
  Commun.}\ }\textbf {\bibinfo {volume} {5}},\ \bibinfo {pages} {3300}
  (\bibinfo {year} {2014})}\BibitemShut {NoStop}%
\bibitem [{\citenamefont {Rodr{\'\i}guez-Fortu{\~n}o}\ \emph
  {et~al.}(2015)\citenamefont {Rodr{\'\i}guez-Fortu{\~n}o}, \citenamefont
  {Mart{\'\i}nez}, \citenamefont {Engheta},\ and\ \citenamefont
  {Zayats}}]{rodriguez2015lateral}%
  \BibitemOpen
  \bibfield  {author} {\bibinfo {author} {\bibfnamefont {F.~J.}\ \bibnamefont
  {Rodr{\'\i}guez-Fortu{\~n}o}}, \bibinfo {author} {\bibfnamefont
  {A.}~\bibnamefont {Mart{\'\i}nez}}, \bibinfo {author} {\bibfnamefont
  {N.}~\bibnamefont {Engheta}}, \ and\ \bibinfo {author} {\bibfnamefont
  {A.~V.}\ \bibnamefont {Zayats}},\ }\href@noop {} {\bibfield  {journal}
  {\bibinfo  {journal} {arXiv preprint arXiv:1504.03464}\ } (\bibinfo {year}
  {2015})}\BibitemShut {NoStop}%
\bibitem [{\citenamefont {Bekshaev}\ \emph {et~al.}(2015)\citenamefont
  {Bekshaev}, \citenamefont {Bliokh},\ and\ \citenamefont
  {Nori}}]{bekshaev2015transverse}%
  \BibitemOpen
  \bibfield  {author} {\bibinfo {author} {\bibfnamefont {A.~Y.}\ \bibnamefont
  {Bekshaev}}, \bibinfo {author} {\bibfnamefont {K.~Y.}\ \bibnamefont
  {Bliokh}}, \ and\ \bibinfo {author} {\bibfnamefont {F.}~\bibnamefont
  {Nori}},\ }\href@noop {} {\bibfield  {journal} {\bibinfo  {journal} {Phys.
  Rev. X}\ }\textbf {\bibinfo {volume} {5}},\ \bibinfo {pages} {011039}
  (\bibinfo {year} {2015})}\BibitemShut {NoStop}%
\bibitem [{\citenamefont {Nieto-Vesperinas}(2015)}]{nieto2015optical}%
  \BibitemOpen
  \bibfield  {author} {\bibinfo {author} {\bibfnamefont {M.}~\bibnamefont
  {Nieto-Vesperinas}},\ }\href@noop {} {\bibfield  {journal} {\bibinfo
  {journal} {arXiv preprint arXiv:1505.05357}\ } (\bibinfo {year}
  {2015})}\BibitemShut {NoStop}%
\bibitem [{\citenamefont {Radescu}\ and\ \citenamefont
  {Vaman}(2002)}]{radescu2002exact}%
  \BibitemOpen
  \bibfield  {author} {\bibinfo {author} {\bibfnamefont {E.}~\bibnamefont
  {Radescu}}\ and\ \bibinfo {author} {\bibfnamefont {G.}~\bibnamefont
  {Vaman}},\ }\href@noop {} {\bibfield  {journal} {\bibinfo  {journal} {Phys.
  Rev. E}\ }\textbf {\bibinfo {volume} {65}},\ \bibinfo {pages} {046609}
  (\bibinfo {year} {2002})}\BibitemShut {NoStop}%
\bibitem [{\citenamefont {Chaumet}\ and\ \citenamefont
  {Rahmani}(2009)}]{chaumet2009electromagnetic}%
  \BibitemOpen
  \bibfield  {author} {\bibinfo {author} {\bibfnamefont {P.~C.}\ \bibnamefont
  {Chaumet}}\ and\ \bibinfo {author} {\bibfnamefont {A.}~\bibnamefont
  {Rahmani}},\ }\href@noop {} {\bibfield  {journal} {\bibinfo  {journal}
  {Optics express}\ }\textbf {\bibinfo {volume} {17}},\ \bibinfo {pages} {2224}
  (\bibinfo {year} {2009})}\BibitemShut {NoStop}%
\bibitem [{\citenamefont {Nieto-Vesperinas}\ \emph {et~al.}(2010)\citenamefont
  {Nieto-Vesperinas}, \citenamefont {S{\'a}enz}, \citenamefont
  {G{\'o}mez-Medina},\ and\ \citenamefont {Chantada}}]{nieto2010optical}%
  \BibitemOpen
  \bibfield  {author} {\bibinfo {author} {\bibfnamefont {M.}~\bibnamefont
  {Nieto-Vesperinas}}, \bibinfo {author} {\bibfnamefont {J.}~\bibnamefont
  {S{\'a}enz}}, \bibinfo {author} {\bibfnamefont {R.}~\bibnamefont
  {G{\'o}mez-Medina}}, \ and\ \bibinfo {author} {\bibfnamefont
  {L.}~\bibnamefont {Chantada}},\ }\href@noop {} {\bibfield  {journal}
  {\bibinfo  {journal} {Opt. Express}\ }\textbf {\bibinfo {volume} {18}},\
  \bibinfo {pages} {11428} (\bibinfo {year} {2010})}\BibitemShut {NoStop}%
\bibitem [{\citenamefont {Chen}\ \emph {et~al.}(2011)\citenamefont {Chen},
  \citenamefont {Ng}, \citenamefont {Lin},\ and\ \citenamefont
  {Chan}}]{chen2011optical}%
  \BibitemOpen
  \bibfield  {author} {\bibinfo {author} {\bibfnamefont {J.}~\bibnamefont
  {Chen}}, \bibinfo {author} {\bibfnamefont {J.}~\bibnamefont {Ng}}, \bibinfo
  {author} {\bibfnamefont {Z.}~\bibnamefont {Lin}}, \ and\ \bibinfo {author}
  {\bibfnamefont {C.}~\bibnamefont {Chan}},\ }\href@noop {} {\bibfield
  {journal} {\bibinfo  {journal} {Nature Photonics}\ }\textbf {\bibinfo
  {volume} {5}},\ \bibinfo {pages} {531} (\bibinfo {year} {2011})}\BibitemShut
  {NoStop}%
\bibitem [{\citenamefont {Bekshaev}(2013)}]{bekshaev2013subwavelength}%
  \BibitemOpen
  \bibfield  {author} {\bibinfo {author} {\bibfnamefont {A.~Y.}\ \bibnamefont
  {Bekshaev}},\ }\href@noop {} {\bibfield  {journal} {\bibinfo  {journal} {J.
  Opt.}\ }\textbf {\bibinfo {volume} {15}},\ \bibinfo {pages} {044004}
  (\bibinfo {year} {2013})}\BibitemShut {NoStop}%
\bibitem [{\citenamefont {Bliokh}\ and\ \citenamefont
  {Nori}(2012)}]{bliokh2012transverse}%
  \BibitemOpen
  \bibfield  {author} {\bibinfo {author} {\bibfnamefont {K.~Y.}\ \bibnamefont
  {Bliokh}}\ and\ \bibinfo {author} {\bibfnamefont {F.}~\bibnamefont {Nori}},\
  }\href@noop {} {\bibfield  {journal} {\bibinfo  {journal} {Phys. Rev. A}\
  }\textbf {\bibinfo {volume} {85}},\ \bibinfo {pages} {061801} (\bibinfo
  {year} {2012})}\BibitemShut {NoStop}%
\bibitem [{\citenamefont {Canaguier-Durand}\ \emph
  {et~al.}(2013{\natexlab{b}})\citenamefont {Canaguier-Durand}, \citenamefont
  {Cuche}, \citenamefont {Genet},\ and\ \citenamefont
  {Ebbesen}}]{canaguier2013force}%
  \BibitemOpen
  \bibfield  {author} {\bibinfo {author} {\bibfnamefont {A.}~\bibnamefont
  {Canaguier-Durand}}, \bibinfo {author} {\bibfnamefont {A.}~\bibnamefont
  {Cuche}}, \bibinfo {author} {\bibfnamefont {C.}~\bibnamefont {Genet}}, \ and\
  \bibinfo {author} {\bibfnamefont {T.~W.}\ \bibnamefont {Ebbesen}},\
  }\href@noop {} {\bibfield  {journal} {\bibinfo  {journal} {Phys. Rev. A}\
  }\textbf {\bibinfo {volume} {88}},\ \bibinfo {pages} {033831} (\bibinfo
  {year} {2013}{\natexlab{b}})}\BibitemShut {NoStop}%
\bibitem [{\citenamefont {Canaguier-Durand}\ and\ \citenamefont
  {Genet}(2014{\natexlab{b}})}]{canaguier2014transverse}%
  \BibitemOpen
  \bibfield  {author} {\bibinfo {author} {\bibfnamefont {A.}~\bibnamefont
  {Canaguier-Durand}}\ and\ \bibinfo {author} {\bibfnamefont {C.}~\bibnamefont
  {Genet}},\ }\href@noop {} {\bibfield  {journal} {\bibinfo  {journal} {Phys.
  Rev. A}\ }\textbf {\bibinfo {volume} {89}},\ \bibinfo {pages} {033841}
  (\bibinfo {year} {2014}{\natexlab{b}})}\BibitemShut {NoStop}%
\bibitem [{\citenamefont {Banzer}\ \emph {et~al.}(2013)\citenamefont {Banzer},
  \citenamefont {Neugebauer}, \citenamefont {Aiello}, \citenamefont
  {Marquardt}, \citenamefont {Lindlein}, \citenamefont {Bauer},\ and\
  \citenamefont {Leuchs}}]{banzer2013photonic}%
  \BibitemOpen
  \bibfield  {author} {\bibinfo {author} {\bibfnamefont {P.}~\bibnamefont
  {Banzer}}, \bibinfo {author} {\bibfnamefont {M.}~\bibnamefont {Neugebauer}},
  \bibinfo {author} {\bibfnamefont {A.}~\bibnamefont {Aiello}}, \bibinfo
  {author} {\bibfnamefont {C.}~\bibnamefont {Marquardt}}, \bibinfo {author}
  {\bibfnamefont {N.}~\bibnamefont {Lindlein}}, \bibinfo {author}
  {\bibfnamefont {T.}~\bibnamefont {Bauer}}, \ and\ \bibinfo {author}
  {\bibfnamefont {G.}~\bibnamefont {Leuchs}},\ }\href@noop {} {\bibfield
  {journal} {\bibinfo  {journal} {J. Eur. Opt. Soc. -Rapid publications}\
  }\textbf {\bibinfo {volume} {8}} (\bibinfo {year} {2013})}\BibitemShut
  {NoStop}%
\bibitem [{\citenamefont {Neugebauer}\ \emph {et~al.}(2014)\citenamefont
  {Neugebauer}, \citenamefont {Banzer}, \citenamefont {Bauer}, \citenamefont
  {Orlov}, \citenamefont {Lindlein}, \citenamefont {Aiello},\ and\
  \citenamefont {Leuchs}}]{neugebauer2014geometric}%
  \BibitemOpen
  \bibfield  {author} {\bibinfo {author} {\bibfnamefont {M.}~\bibnamefont
  {Neugebauer}}, \bibinfo {author} {\bibfnamefont {P.}~\bibnamefont {Banzer}},
  \bibinfo {author} {\bibfnamefont {T.}~\bibnamefont {Bauer}}, \bibinfo
  {author} {\bibfnamefont {S.}~\bibnamefont {Orlov}}, \bibinfo {author}
  {\bibfnamefont {N.}~\bibnamefont {Lindlein}}, \bibinfo {author}
  {\bibfnamefont {A.}~\bibnamefont {Aiello}}, \ and\ \bibinfo {author}
  {\bibfnamefont {G.}~\bibnamefont {Leuchs}},\ }\href@noop {} {\bibfield
  {journal} {\bibinfo  {journal} {Physical Review A}\ }\textbf {\bibinfo
  {volume} {89}},\ \bibinfo {pages} {013840} (\bibinfo {year}
  {2014})}\BibitemShut {NoStop}%
\bibitem [{\citenamefont {Neugebauer}\ \emph {et~al.}(2015)\citenamefont
  {Neugebauer}, \citenamefont {Bauer}, \citenamefont {Aiello},\ and\
  \citenamefont {Banzer}}]{neugebauer2015measuring}%
  \BibitemOpen
  \bibfield  {author} {\bibinfo {author} {\bibfnamefont {M.}~\bibnamefont
  {Neugebauer}}, \bibinfo {author} {\bibfnamefont {T.}~\bibnamefont {Bauer}},
  \bibinfo {author} {\bibfnamefont {A.}~\bibnamefont {Aiello}}, \ and\ \bibinfo
  {author} {\bibfnamefont {P.}~\bibnamefont {Banzer}},\ }\href@noop {}
  {\bibfield  {journal} {\bibinfo  {journal} {Phys. Rev. Lett.}\ }\textbf
  {\bibinfo {volume} {114}},\ \bibinfo {pages} {063901} (\bibinfo {year}
  {2015})}\BibitemShut {NoStop}%
\bibitem [{\citenamefont {Aiello}\ and\ \citenamefont
  {Banzer}(2015)}]{aiello2015transverse}%
  \BibitemOpen
  \bibfield  {author} {\bibinfo {author} {\bibfnamefont {A.}~\bibnamefont
  {Aiello}}\ and\ \bibinfo {author} {\bibfnamefont {P.}~\bibnamefont
  {Banzer}},\ }\href@noop {} {\bibfield  {journal} {\bibinfo  {journal} {arXiv
  preprint arXiv:1502.05350}\ } (\bibinfo {year} {2015})}\BibitemShut {NoStop}%
\end{thebibliography}%

\end{document}